\documentclass[prd,aps,preprint,showpacs,showkeys,amsmath,amssymb,preprintnumbers,superscriptaddress,floatfix]{revtex4-1}%

\usepackage{graphicx}
\usepackage{epstopdf}

\usepackage{array,amsmath,amssymb}
\usepackage{amsmath}
\usepackage{amssymb}

\newcommand{\Z}{{\mathbb{Z}}}

\def\vc#1{ \boldsymbol #1}

\begin{document}

\title{Gluon propagators and center vortices in gluon plasma} 

\author{M.N.~Chernodub}\thanks{On leave from
 Institute of Theoretical and Experimental Physics, Moscow, Russia.}
\affiliation{CNRS, Laboratoire de Math\'ematiques et Physique Th\'eorique,
Universit\'e Fran\c{c}ois-Rabelais, \\ F\'ed\'eration Denis Poisson - CNRS,
Parc de Grandmont, Universit\'e de Tours, 37200 France}
\affiliation{Department of Physics and Astronomy, University of Gent, Krijgslaan 281, S9, B-9000 Gent, Belgium}

\author{Y.~Nakagawa}
\affiliation{Graduate School of Science and Technology, Niigata University,
 Niigata 950-2181, Japan}

\author{A.~Nakamura}
\affiliation{Research Institute for Information Science and Education,
Hiroshima University, Higashi-Hiroshima 739-8521, Japan}

\author{T.~Saito}
\affiliation{Integrated Information Center, Kochi University, Kochi, 780--8520, Japan}

\author{V.~I.~Zakharov}
\affiliation{Institute of Theoretical and Experimental Physics, 117259, Moscow, Russia}
\affiliation{Max-Planck Institut f\"ur Physik, F\"ohringer Ring 6, 80805, M\"unchen, Germany}

\begin{abstract}
We study electric and magnetic components of the gluon propagators 
in quark-gluon plasma in terms of center vortices 
by using a quenched simulation of $SU(2)$ lattice theory.
In the Landau gauge, the magnetic components of the propagators  
are strongly affected  in the infrared region by removal of the center vortices, 
while the electric components are almost unchanged by this procedure. In the Coulomb gauge,
the time-time correlators, including an instantaneous interaction, also have an essential contribution from the center vortices.
As a result, one finds that magnetic degrees of freedom in the infrared region
couple strongly to the center vortices in the deconfinement phase.
\end{abstract}


\maketitle

\section{Introduction}

The Relativistic Heavy-Ion Collider (RHIC) at Brookhaven 
 National Laboratory produces a new state of matter which may exceed the 
critical temperature $T_c$ of the phase transition
from the hadron phase to the quark-gluon plasma (QGP) phase.
Many phenomenological studies and lattice computations 
suggest that the QGP is a strongly interacting plasma \cite{Shu1},
for which we cannot apply the early   
arguments based on the perturbative approach with a small coupling constant.
Furthermore, the recent Pb-Pb heavy-ion collision experiment 
at LHC has created QGP matter 
at even higher temperatures: This shows us an obvious
 jet-quenching event \cite{LHC1} and a larger elliptic flow \cite{LHC2}
 compared to the RHIC's Au-Au collision.
Therefore, it is indispensable for us to explore 
the mechanism which drives such strong interactions
using a nonperturbative first-principle approach in lattice simulations.

One of the most important ideas to describe a strongly interacting QGP (sQGP)  
is to focus on an infrared 
singularity arising from magnetic degrees of freedom \cite{Linde,TQCD}.
The magnetic component of the gluon propagator is fully inaccessible by the 
perturbative calculation, but its infrared divergence
may cause an emergence of a nonperturbative magnetic mass that plays a cutoff role 
and can cure thermal QCD in the infrared region. 
The lattice simulations \cite{Heller,ATS, Maezawa}
prove that the magnetic gluons have a nonvanishing mass at finite temperature. 
Furthermore, it is well known that a spatial Euclidean Wilson loop 
(which is not extended to the temporal dimension) bears a confining potential above $T_c$
 \cite{SWL1,SWL2,SWL3,SWL4,SWL5,SWL6,SWL7,SWL8,SWL9}, 
while the correlators of a Polyakov line -- wrapped in the temporal direction --
give a nonconfining screened potential of the Debye type,  
with a finite electric mass $\sim g(T)T$ \cite{CDP1,CDP2,SWL9,Kacz}.

In addition, from the viewpoint of the Gribov-Zwanziger (GZ) confinement scenario \cite{Gribov,CCG,Zwan},
a color-Coulomb instantaneous interaction between a quark and an antiquark
provides -- even in the nonconfined QGP phase -- a confining potential which rises linearly as the function of the quark-antiquark separation~\cite{Green1,Green2,Sai1,Sai2}. As a consequence, the thermal string tensions obtained from the spatial-Wilson and the color-Coulomb potentials are nonzero. They depend on the temperature and obey a magnetic scaling law [$\sim g^2(T)T$]. Extending this line of considerations, Zwanziger has 
approximately 
reconstructed the equation of state of QGP using the Gribov-type dispersion relation for the massive gluons~\cite{EOSfFMR}.

These interesting aspects of the non-Abelian gauge theory may be related to center (magnetic) vortices -- i.e., to the topological defects associated with
 the nontrivial homotopy group $\pi_1[SU(N)/\Z(N)]\sim \Z(N)$ -- which are responsible for 
certain
nonperturbative phenomena of QCD. 
One can identify the center vortices on the lattice using 
a numerical technique~\cite{DFGGO} 
and also remove these vortices from the original gauge fields \cite{FD}. It turns out that the removal of the center vortices destroys the color confinement property and restores the chiral symmetry.
Moreover, the lattice center-vortex density exhibits a scaling consistent with the asymptotic freedom~\cite{Lang1}.

In terms of the vortex degrees of freedom, the QCD deconfinement phase transition can be considered as a depercolation transition of the vortex lines in the direction of the Euclidean time~\cite{Lang2}. As a result, we can naturally understand the 
survival
of the spatial confinement above $T_c$ because the center vortices remain intact in the spatial space.
Moreover, 
a typical center-vortex configuration 
is located at the Gribov horizon
in the gauge space.
Thus, the removal of the center vortices results
 in the dilution of the lowest eigenvalues of the Faddeev-Popov operator.
These eigenvalues -- according to the GZ confinement scenario --
 cause confinement of color~\cite{Rein,Green2}.

Recently, three of us have argued that the center-vortex mechanism is also important in the hot phase of the Yang-Mills theory because the center vortices carry information about  the magnetic degrees of freedom~\cite{CV,CNV}. The center vortices are related to Abelian magnetic monopoles, and the later are expected to explain some of the interesting properties of the quark-gluon plasma as well~\cite{Shu2}.

In this paper we study a connection between the center vortices and the infrared properties of the gluon propagators at finite temperature. To this end we study the behavior of the electric and magnetic components of the gluon propagators by removing the vortices from the original gauge configurations and comparing the result with the original one.  We use the quenched $SU(2)$ lattice simulations in the Landau and Coulomb gauges. In Sec. II we define gluon propagators on the lattice. In Sec. III a numerical technique
 used to make a center projection is summarized.
Our numerical results are presented in the Sec. IV, while the last section is devoted to the summary of this work.

\section{Gluon propagators} 

In this study, we work in the $SU(2)$ lattice gauge theory. The gauge potential $A$ is expressed via the $SU(2)$ matrix link variable $U_{\mu}(x)$ as follows:
\begin{equation}
A_{\mu}(\vc{x},t) = \frac{1}{2} \sum_a \mbox{Tr} \, \sigma^a U_{\mu}(\vc{x},t)\,,
\label{gf}
\end{equation}
where $\sigma^{a}$ are the Pauli matrices.
The correlation
functions of the gauge fields~(\ref{gf}) in momentum space are
\begin{equation}
D_{\mu\nu}(\vc{q},t) = \frac{1}{3V}
 \sum_{\vc{x},\vc{y}}
 \langle A_{\mu}(\vc{x},t^{\prime}) A_{\nu}(\vc{y},t^{\prime\prime})
 \rangle
 e^{i \vc{q}(\vc{x}-\vc{y})},\label{cinm} 
\end{equation}
where $V$($=N_xN_yN_z$) is the three-dimensional volume and $t=t^\prime-t^{\prime\prime}$ is the Euclidean time difference.

In Landau-gauge fixing we study the static  correlators of gluon fields with $q_0 = 0$:
\begin{equation}
D_{\mu\nu}(\vc{q},q_0=0) = \frac{1}{N_t} \sum_t D_{\mu\nu}(\vc{q},t)\,,
\label{cinm2}
\end{equation}
where $N_t$ is the lattice size in the Euclidean temporal direction.
In the Coulomb gauge, it is more appropriate to investigate an equal-time gluon propagator in the following form:
\begin{equation}
D_{\mu\nu}^{eq}(\vc{q}) = \frac{1}{3VN_t}
\sum_{\vc{x},\vc{y},t} \langle A_{\mu}(\vc{x},t) A_{\nu}(\vc{y},t) \rangle
 e^{i \vc{q}(\vc{x}-\vc{y})}. 
\label{cinm3}
\end{equation}
This propagator corresponds to the one in 
Eq. (\ref{cinm}) at $t^{\prime}=t^{\prime\prime}$.
Note that there is no $q_0$ dependence 
in Eq. (\ref{cinm}) and that 
the $q_0=0$ term is removed from the sum.
An equal-time propagator reads
$D(q)=1/(2\omega(\vc{q}))$ where  $\omega=\sqrt{\vc{q}^2+m^2}$ 
is the dispersion relation.

In the finite-temperature system, 
the electric and magnetic gluons have different effects due to breaking of the Euclidean Lorentz invariance.
One can define the spatially transverse ($P_T$) and spatially longitudinal ($P_L$) projection operators as follows:
\begin{eqnarray}
P_T^{00}=P_T^{0i}=P_T^{i0},\hspace{0.5cm} 
P_T^{ij}=\delta^{ij}-\frac{q^i q^j}{{q_i^2}}, \\
P_L^{\mu\nu}=\delta^{\mu\nu}-\frac{q^{\mu} q^{\nu}}{{{q}}^2} - P_T^{\mu\nu}\,, 
\end{eqnarray}
with the properties
\begin{eqnarray}
(P_T)^2 = P_T,\hspace{0.5cm}(P_L)^2 = P_L,\hspace{0.5cm}P_T P_L = 0\,.
\end{eqnarray}
Both spatially transverse and spatially longitudinal projectors correspond to
 the transverse states in momentum space:
\begin{equation}
q_\mu P_T^{\mu\nu} =  q_\mu P_L^{\mu\nu}= 0\,.
\end{equation}
Using these relations, the gluon propagators at finite temperature 
in a Landau-type gauge can be separated into two independent parts: 
\begin{equation}
D_{\mu\nu} = \frac{1}{G+q^2}  P_{T}^{\mu\nu} + \frac{1}{F+q^2}  P_{L}^{\mu\nu} \,. 
\label{gp}
\end{equation}
The electric component of the gluon propagator is given by the spatially longitudinal projection,
$D_e = D_{00}$ and the electric mass is given by
 $F((\vc{q},q_0)=0)=m_e^2\sim (g(T)T)^2$.
The spatially transverse projection gives us the magnetic propagator
 $D_m = D_{ii}$. 
The magnetic mass is expected to be $G((\vc{q},q_0)=0)=m_m^2 \sim (g^2(T)T)^2$,
 where $g(T)$ is a running QCD coupling defined at the scale of temperature $T$.

\section{Maximal center projection}

We employ a direct maximal center projection (MCP)~\cite{DFGGO} in order to identify the center vortices on the lattice. The corresponding gauge is defined by the condition
\begin{equation}
\mbox{Maximize}\hspace{0.1cm} R = \frac{1}{VN_t} \sum_{x} \mbox{Tr}
\left[ U_{\mu}(x) \right]^2\,. \label{mcpe}
\end{equation}
The center gauge field,
\begin{equation}
Z_{\mu}(x)=\mbox{sgn} \mbox{Tr} \left[ U_{\mu}(x) \right]  \in \mathbb{Z}_2\,,
\end{equation}
allows us to identify the center vortices. If the center plaquette is not equal to a trivial element (unity) then a center vortex goes through this plaquette.

In order to remove the center vortices from the gauge-field ensemble,
 we follow Ref.~\cite{FD} by 
multiplying the original field $U_{\mu}$
 by the center-projected field $Z_{\mu}$:
\begin{equation}
U_{\mu}^{'}(x) = Z_{\mu} U_{\mu}(x)\,,
\end{equation}
so that the new links $U_{\mu}^{'}$ correspond to vortex-free ensembles.

It is confirmed by lattice simulations that 
the confinement and chiral symmetry breaking
are both lost after the removal of the center vortices~\cite{DFGGO,FD}.
 We would also like to note that the effect of the vortex removal on chiral symmetry breaking ($\chi$SB) depends on the choice of the lattice quark action~\cite{Gattnar05,Faber08,Bowman08}, and thus chiral symmetry breaking should be treated with care.
In our paper we address the problem of color confinement.

\section{Numerical results}

\subsection{Lattice setup}
We carry out quenched $SU(2)$ lattice simulations by generating gauge
 configurations.
The convergence criterion of the MCP technique is set as $10^{-16}$,
 and the precision of an iterative gauge fixing algorithm \cite{Mandula} 
is $10^{-8}$.
We use the single geometry of the lattice, $24^3\times4$,
 and various temperatures $T/T_c \sim 1.4$, $3.0$, and $6.0$
 [$T_c \sim 305$\,MeV for $N_t=4$ for the $SU(2)$ gauge group] for $\beta=2.40$, $2.64$, and $2.88$, respectively \cite{CNV,Kar}. We used approximately $20$ to $30$ lattice configurations collected every 100 sweep steps.

\subsection{Thermal gluon propagators}

\begin{figure*}[htbp]
\begin{center}
\begin{tabular}{ccc}
\includegraphics[width=60mm,clip=true]{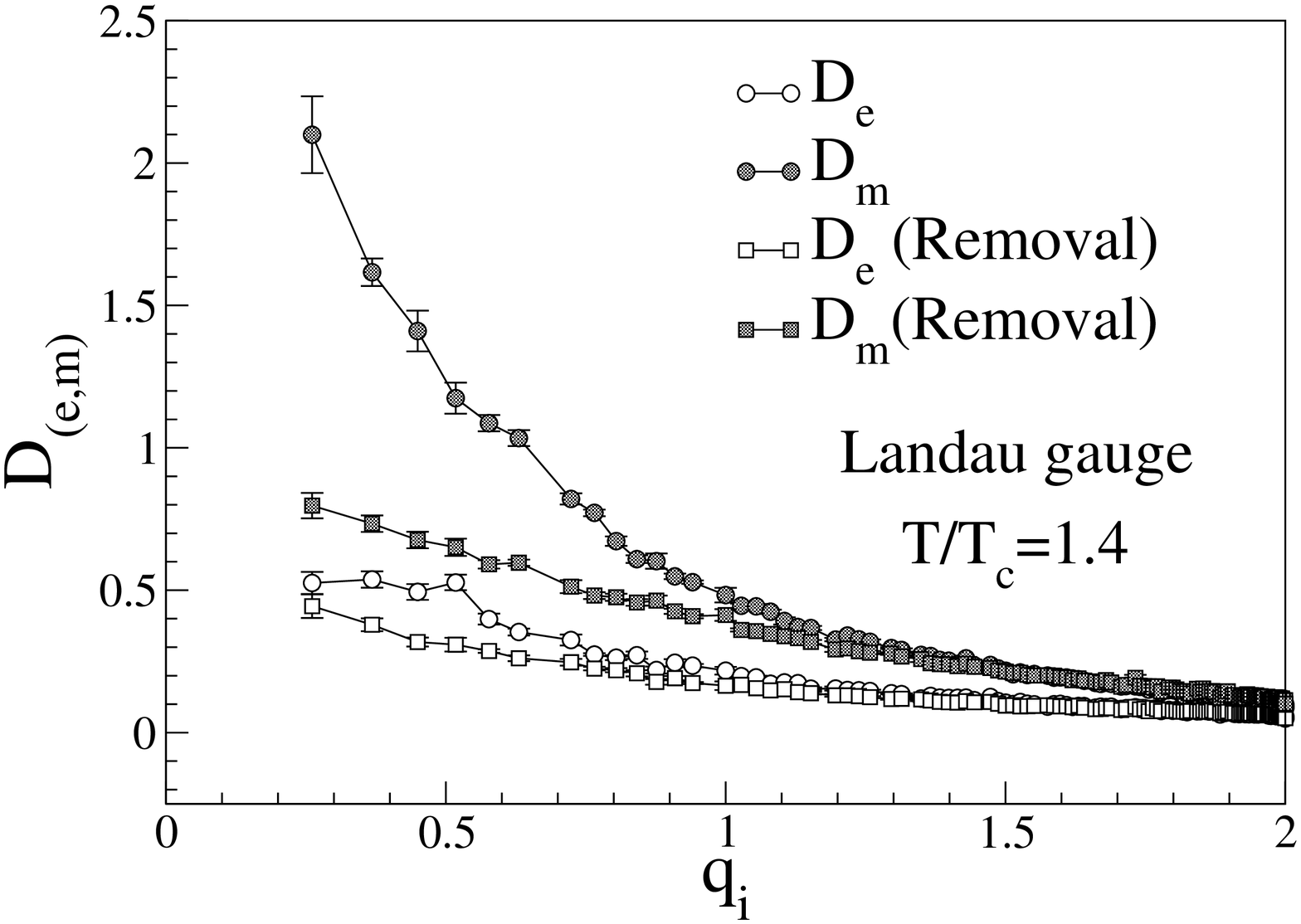} & 
\includegraphics[width=60mm,clip=true]{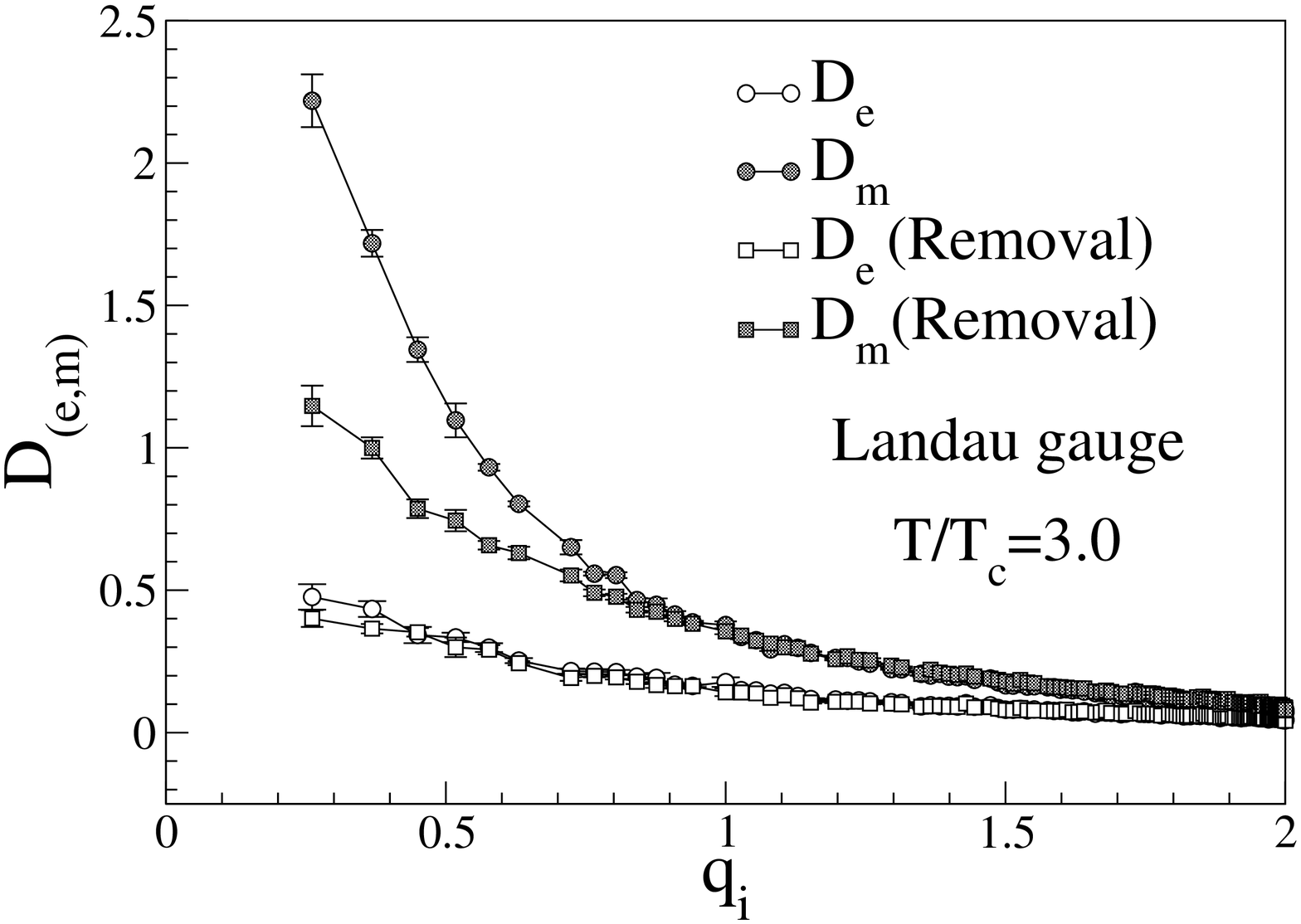} &
\includegraphics[width=60mm,clip=true]{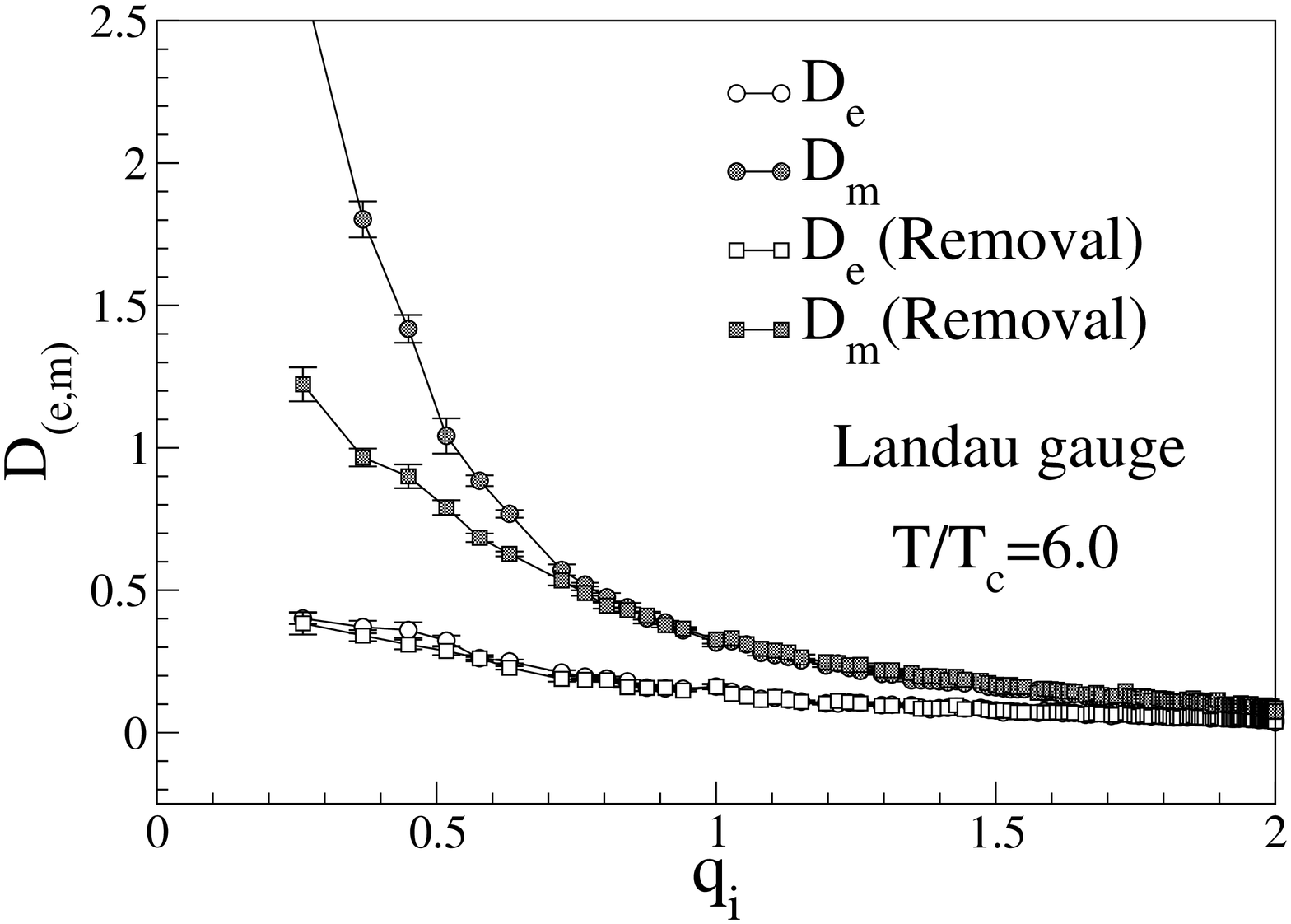}
\end{tabular}
\caption{Electric and magnetic gluon propagators 
in the Landau gauge for $T/T_c= 1.4$, $3.0$, and $6.0$
as a function of the spatial momentum.
The circle symbols represent numerical results obtained with the original lattice configurations, while the square symbols 
correspond to the vortex-removed configurations. 
}
\label{G1}
\end{center}
\end{figure*}

In the left plot of Fig. \ref{G1} we show the gluon propagators calculated in the Landau gauge at $T/T_c=1.40$. The removal of center vortices visibly affects the infrared behavior of the magnetic and electric gluons in the infrared region. However, the effect of 
the vortex removal is much more pronounced for 
the magnetic degrees of freedom
compared to the effect on the electric correlators.

The effect of the vortex removal on the electric component of the gluon propagator diminishes with an increase of temperature, as one can see from the plots of Fig.~\ref{G1}, corresponding to the higher temperatures $T/T_c=3.0$ and $6.0$. However, the magnetic propagators are affected drastically by the center vortices in the infrared region for all studied temperatures.
 
\begin{figure*}[htbp]
\begin{center}
\begin{tabular}{ccc}
\includegraphics[width=60mm,clip=true]{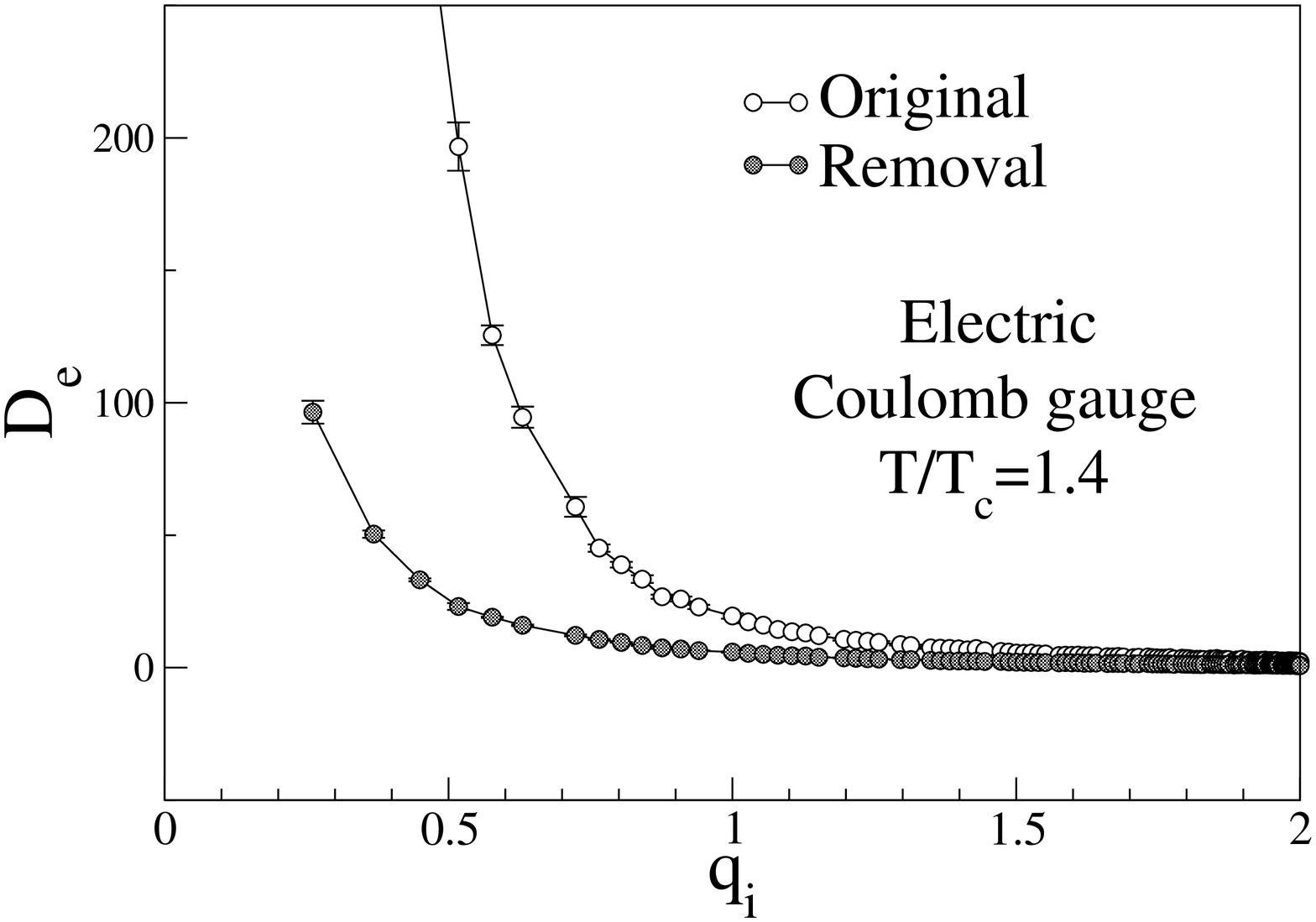}
\includegraphics[width=60mm,clip=true]{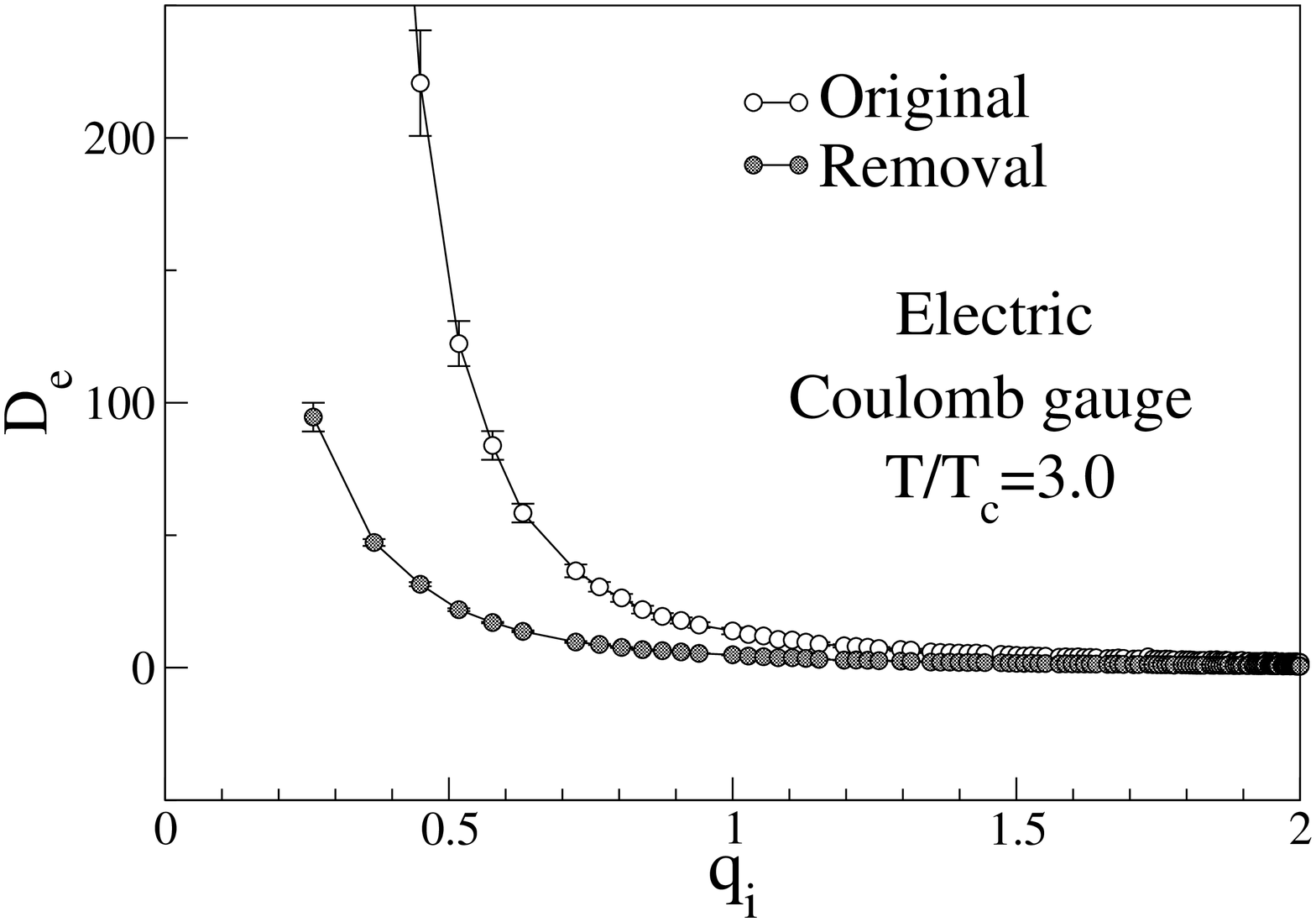}
\includegraphics[width=60mm,clip=true]{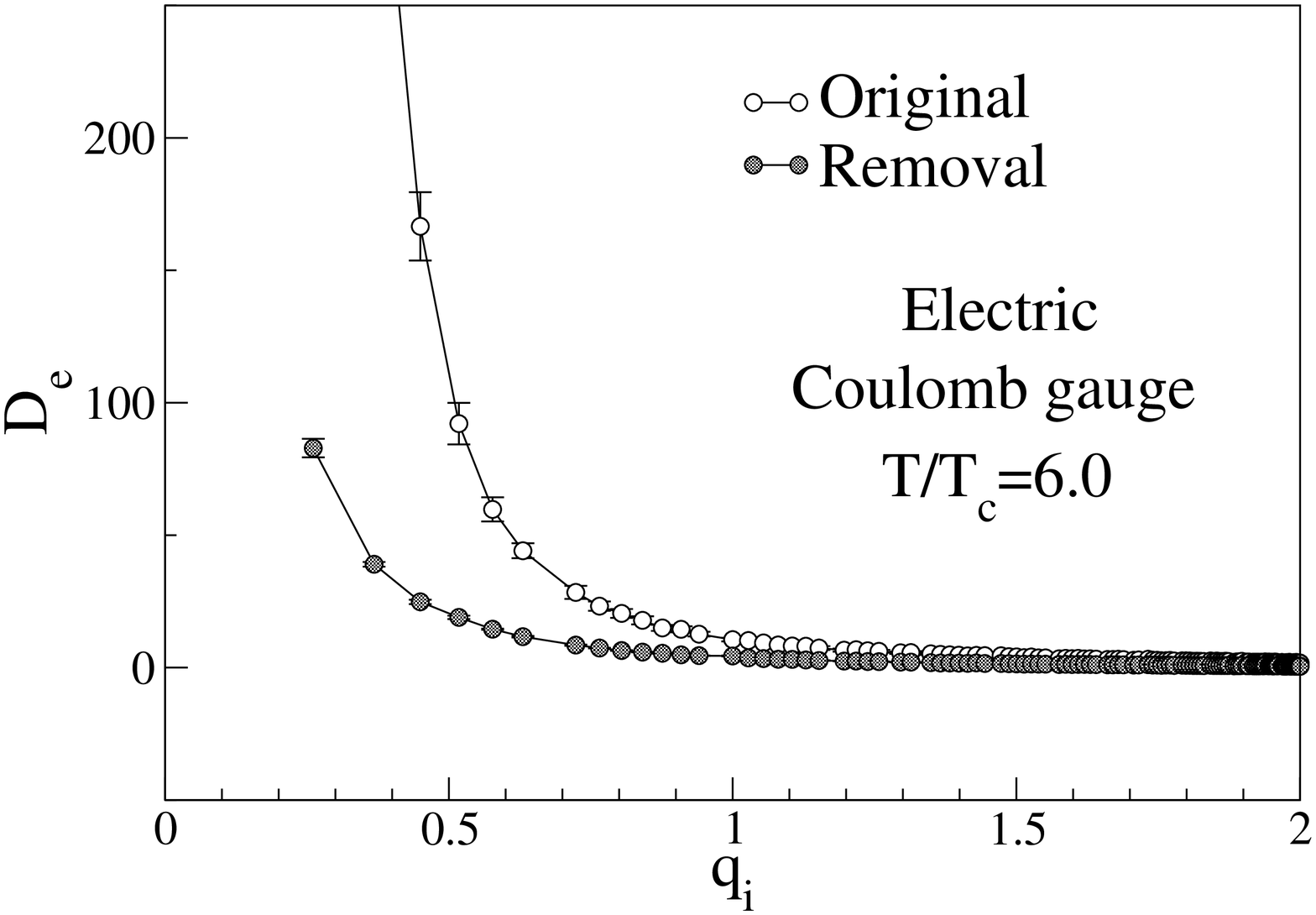} \\[-6mm]
\includegraphics[width=60mm,clip=true]{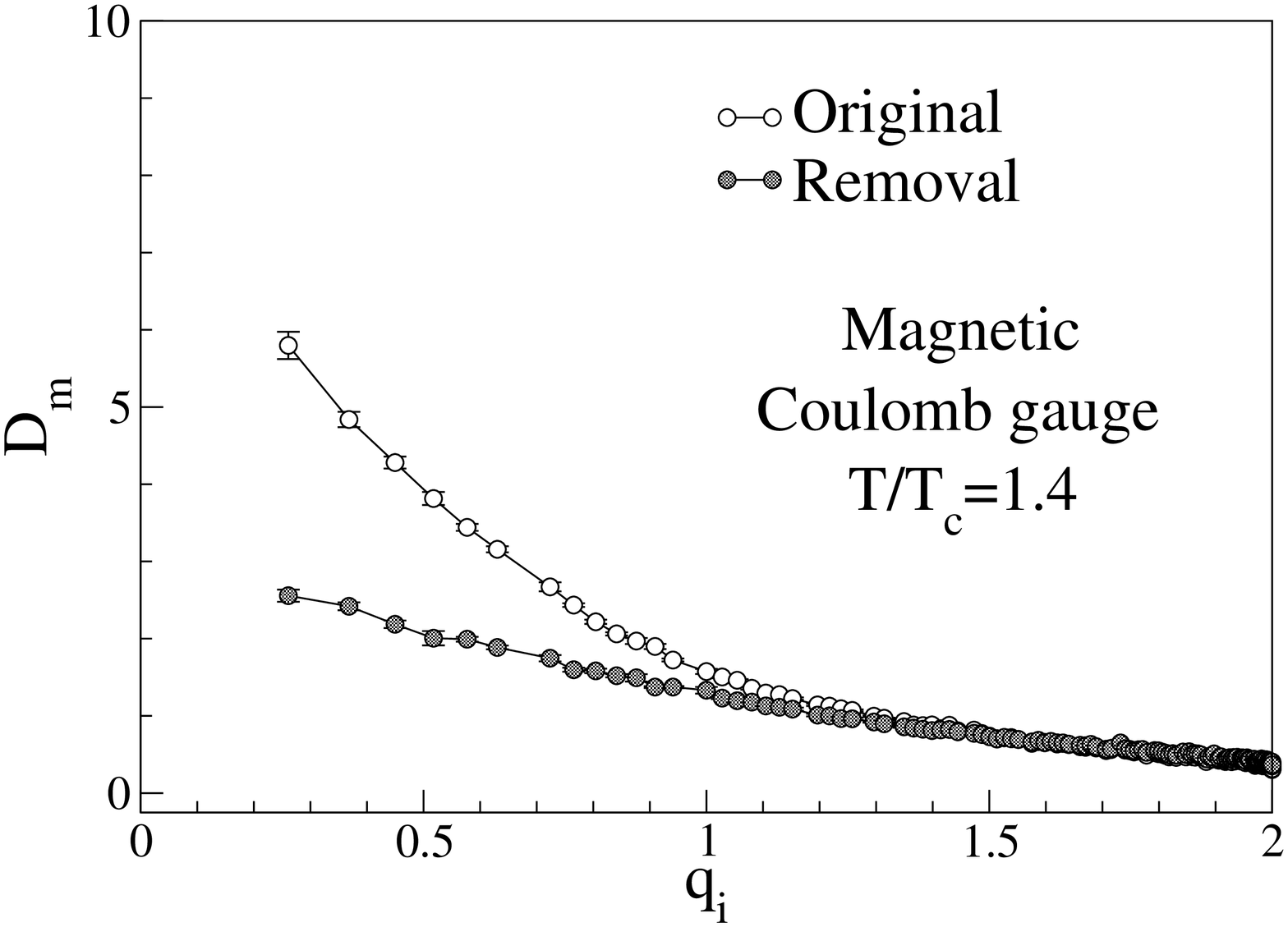}
\includegraphics[width=60mm,clip=true]{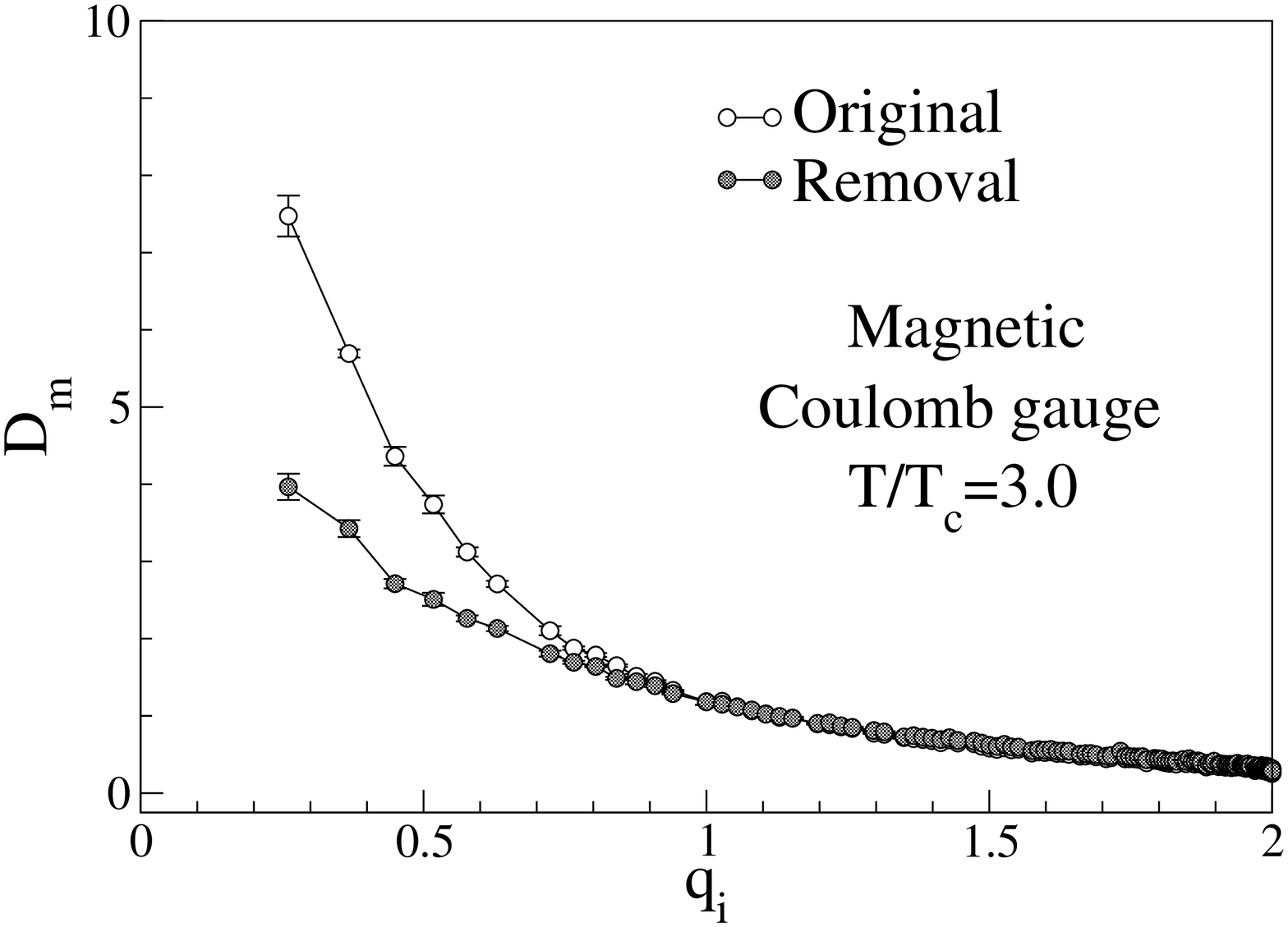}
\includegraphics[width=60mm,clip=true]{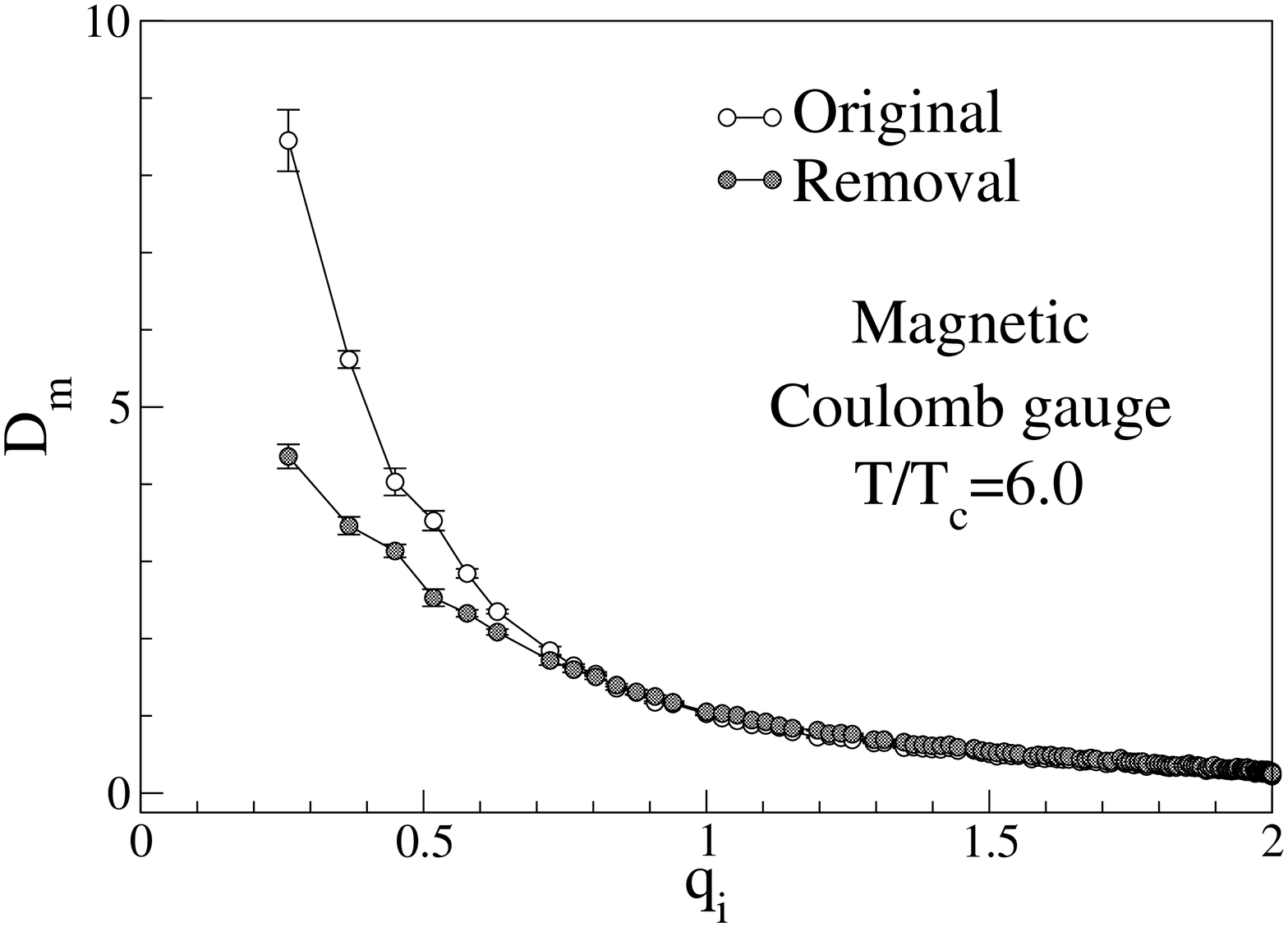}
\end{tabular}
\caption{Electric (upper figures) and magnetic (lower figures)
 gluon propagators 
in the Coulomb gauge as a function of the spatial momentum at $T/T_c= 1.4$, $3.0$ and $6.0$.
The open symbols represent the numerical results obtained with the use of the original lattice configurations, 
while the filled symbols correspond to the vortex-removed configurations.}
\label{G2}
\end{center}
\end{figure*}

The Coulomb gauge gluon propagators are plotted in Fig. \ref{G2}.
The magnetic propagator is affected by 
the removal of the center vortices
in the deconfinement regions,
being consistent with that of the Landau-gauge case 
as we discussed in the previous paragraph.
Contrary to this fact, the electric parts in the Coulomb gauge are
 influenced by the magnetic vortices.
This tendency remains for $T/T_c = 1 \sim 6$; thus, it seems that 
there is an inconsistency between the two gauges. 

However, we have to mention the fact that in the Coulomb gauge the  temporal-gauge correlator is dominated by an infrared singularity arising from spatial (magnetic) components~\cite{CCG,Zwan,Green1,Green2,Sai1,Sai2,Sai3,Sai4,Sai5}.
The time-time correlator with Coulomb-gauge fixing can be decomposed into two parts:
\begin{equation}
D_{00}(\vc{x},t) = V_c(\vc{x})\delta(t) + P(\vc{x},t),
\end{equation}
where $V_c$ is an instantaneous potential, which is responsible for 
the color confinement and $P$ corresponds to the vacuum (retarded) polarization term.
In this theory, $V_c$ is related to
the Green's function $M^{-1}$ of the Faddeev-Popov ghost,
\begin{equation} 
V_c(\vc{x}-\vc{y}) \delta_{ab} = \left<
(M^{-1}(-\partial^2_i)M^{-1})_{\vc{x},\vc{y}}^{ab}\right>,
\end{equation}
which does not explicitly depend on the Euclidean time (temperature), 
and thus, this quantity has no effect on screening.
In contrast, $P$
 is a function of time and it may contribute to the screening.
Indeed, in the deconfinement phase,  this term provides a screened quark potential with finite electric mass.
The screening can be observed by the investigation of 
a Polyakov line correlator \cite{Maezawa}.
The Polyakov line correlator with Landau-gauge fixing 
gives the color-screened potential as well \cite{CDP1,CDP2}.

In the confinement region, $V_c$ is a linearly rising potential.
Moreover, even above the critical temperature $T_c$ the potential $V_c$ is a confining potential \cite{Green2}.
Its thermal color-Coulomb string tension depends on temperature.
The temperature dependence is consistent with the magnetic scaling $g^2(T)T$ \cite{Sai2}.
The remnant confinement property corresponds to 
the nonvanishing spatial string tension. 
Consequently, it is now obvious that the time correlator in the
Coulomb gauge is also a magneticlike quantity.

Conversely, the covariant-type Landau gauge may not plainly separate the longitudinal and transverse modes.
Actually, it is more difficult to observe a confining property of gluons 
even in the confinement region, compared to the case of the Coulomb gauge \cite{Cucc}.
However, as seen in Fig.~\ref{G1}, the physical magnetic gluon 
is definitely affected by the vortex removal,
 while the corresponding variation of the electric gluon 
(excluding the instantaneous interaction which is singular in the infrared region)
is very small.

It is natural that the confining behavior of the thermal gluon
propagators has different forms for different gauge fixings.
Nevertheless, 
it is very important to stress that the calculations 
in both gauges
give us the same conclusion that the relevant elements
 to the magnetic degrees of freedom are strongly coupled
 to center vortices after the deconfining phase transition. 

In addition, our result means that the Gribov-Zwanziger confinement scenario survives above $T_c$.
According to this theory, the spatial correlator 
experiences the suppression effects in the confinement phase,
 and the temporal correlator diverges 
in the infrared limit. We see that a similar behavior is seen
 in our numerical data at finite temperature.
Furthermore, our observation can also be derived from the fact that  in the QGP phase the vortex configurations belong
to the Gribov horizon~\cite{Green2}.

\subsection{Check for numerical ambiguities}

\begin{figure*}[htbp]
\begin{center}
\begin{tabular}{cc}
\includegraphics[width=90mm,clip=true]{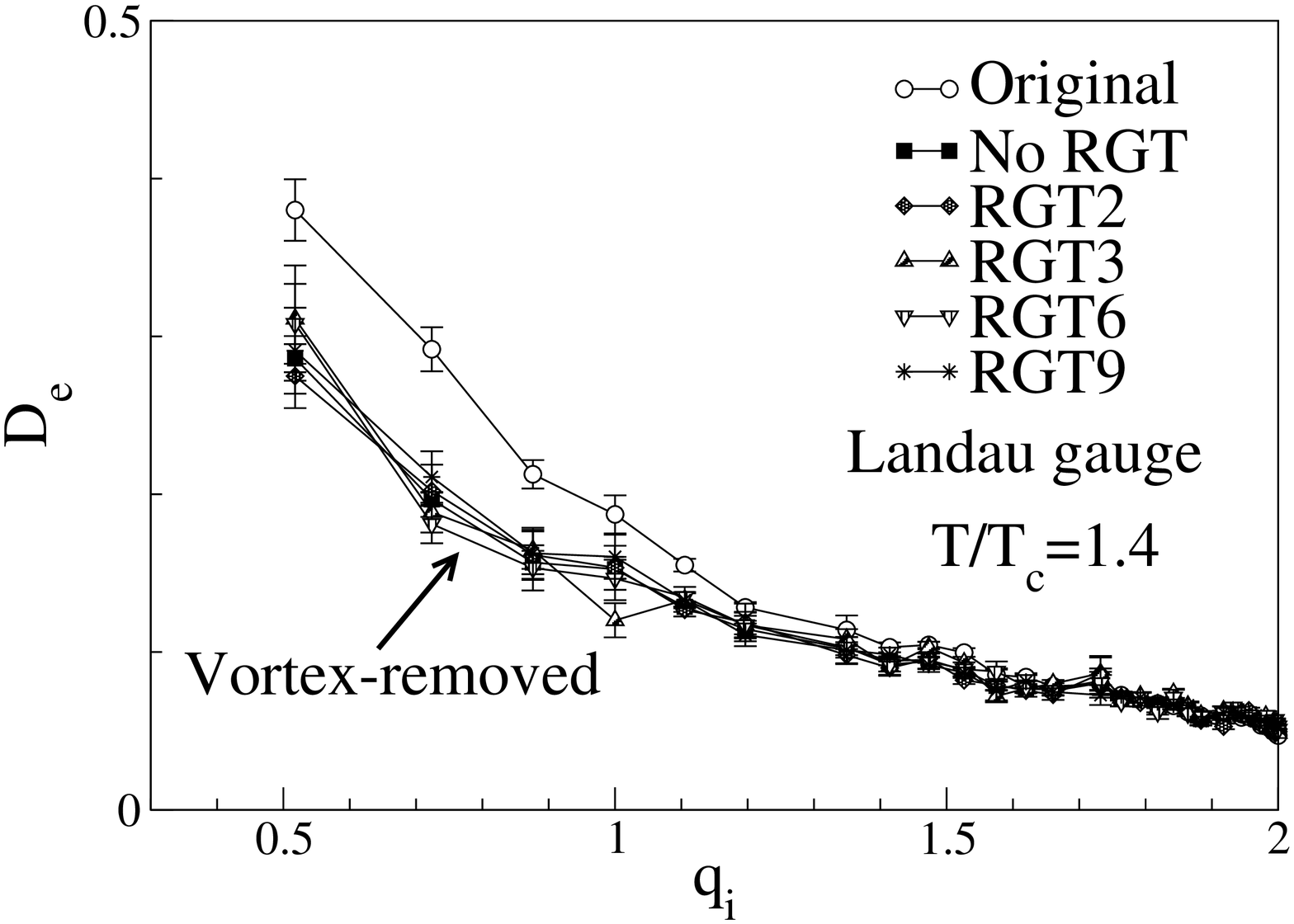} &
\includegraphics[width=90mm,clip=true]{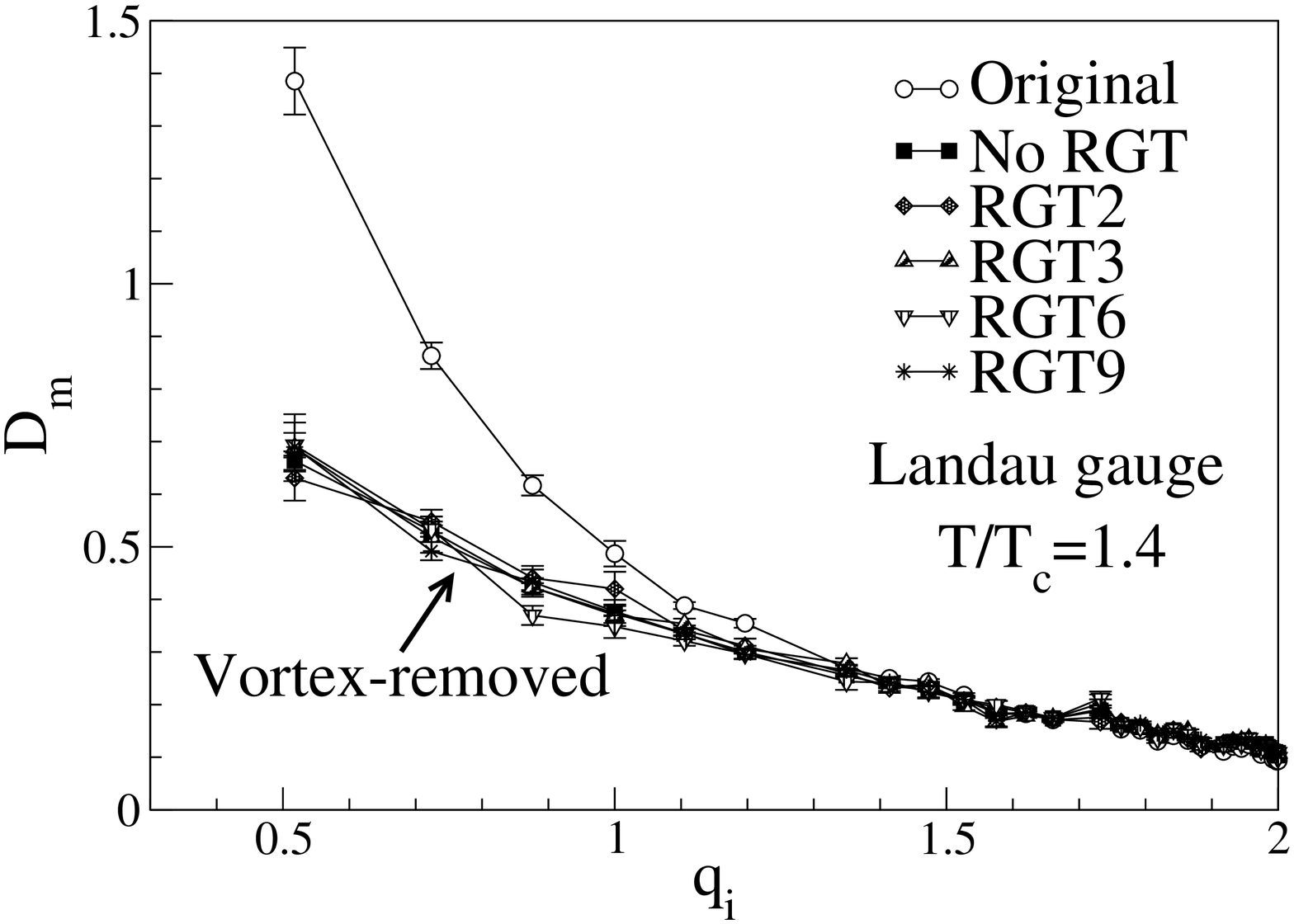}
\end{tabular}
\caption{The dependence of the electric (left panel)
 and magnetic (right panel) gluon propagators 
on the Gribov copy ambiguities.
 The lattice size is $24^3 \times 4$ and $T/T_c =1.4$. 
The RGT with different random seeds is applied to
 each lattice configuration a varying number of times.}
\label{G3}
\end{center}
\end{figure*}

It is well known that the MCP has numerical ambiguities (the Gribov copies). 
In order to check the stability of our results against this ambiguity,
 we used random-gauge transformations (RGT)
applied to the Monte-Carlo updated gauge-field configurations before performing the MCP.
Although the global maximum of the gauge fixing functional~(\ref{mcpe}) cannot be determined with an ideal accuracy,
neither the electric nor the magnetic gluon propagator significantly
 depends on this algorithm, as is shown in Fig.~\ref{G3}.

\section{Summary}

We have studied the effects of the center (magnetic) vortices on electric and magnetic components of the gluon propagators in the QGP phase using $SU(2)$ lattice simulations in the Landau and Coulomb gauges. We find that the gluon dynamics in the infrared region strongly couple to the magnetic vortices. Thus, in the deconfinement phase the magnetic vortex degrees of freedom should be treated nonperturbatively.

At high temperatures the removal of the magnetic vortices reduces drastically the magnetic gluon propagators in the infrared region in both gauges. 
The effect is similar to the suppression of the infrared gluon propagators 
in the confinement phase, observed first in Ref.~\cite{GluonConfinement}. The electric propagators are almost unaffected by this procedure in the Landau gauge while in the Coulomb gauge -- in agreement with the GZ mechanism -- the electric gluon propagator is suppressed by the removal of the center vortices. 

\section{Acknowledgments} 
The authors are grateful to J.~Greensite for useful discussions.
The simulation was performed on SX-8 and SX-9 (NEC) vector-parallel computers 
at the RCNP and the CMC of Osaka University.
 The work of MNC was partially supported by Grant No. ANR-10-JCJC-0408 HYPERMAG.
The work of VIZ is partially supported by Leading Scientific 
Schools Grants No. NSh-6260.2010.2, No. RFBR-11-02-01227,
 and by the Federal Special-Purpose Programme ``Cadres'' 
of the Russian Ministry of Science and Education.

\end{document}